\documentclass[aps,prd,twocolumn,english,nofootinbib]{revtex4-1}
\usepackage[T1]{fontenc}
\usepackage{circuitikz}
\usepackage{makecell}

\usetikzlibrary{shapes.geometric,arrows}

\tikzstyle{startstop} = [cylinder, rounded corners, minimum width=1.5cm, minimum height=0.5cm, text width=0.8cm, text centered, draw=black, fill=blue!30, shape border rotate=90]
\tikzstyle{io} = [trapezium, trapezium left angle=70, trapezium right angle=110, minimum width=4cm, minimum height=1cm, text width=2.5cm, text centered, draw=black, fill=blue!30]
\tikzstyle{process} = [rectangle, text width=3cm, minimum width=3cm, minimum height=1cm, text centered, draw=black, fill=yellow!30]
\tikzstyle{processl} = [rectangle, text width=3cm, minimum width=3cm, minimum height=1cm, text centered, draw=black, fill=green!30]
\tikzstyle{postprocess} = [rectangle, text width=3.8cm, minimum width=4cm, minimum height=1cm, text centered, draw=black, fill=yellow!30]
\tikzstyle{postprocessl} = [rectangle, text width=3.8cm, minimum width=4cm, minimum height=1cm, text centered, draw=black, fill=green!30]
\tikzstyle{exprocess} = [rectangle, text width=3cm, minimum width=3cm, minimum height=1cm, text centered, draw=black, fill=red!30]
\tikzstyle{exprocessl} = [rectangle, text width=3.8cm, minimum width=4cm, minimum height=1cm, text centered, draw=black, fill=red!30]
\tikzstyle{inprocess} = [rectangle, text width=2.5cm, minimum width=3cm, minimum height=1cm, text centered, draw=black, fill=green!30]
\tikzstyle{decision} = [diamond, minimum width=1.5cm, minimum height=0.7cm, text width=2.08cm, text centered, draw=black, fill=green!30]
\tikzstyle{arrow} = [thick, ->, >=stealth]
\tikzstyle{arrowboth} = [thick, <->, >=stealth]

\usepackage[unicode=true,pdfusetitle,
 bookmarks=true,bookmarksnumbered=false,bookmarksopen=false,
 breaklinks=false,pdfborder={0 0 1},backref=false,colorlinks=true]{hyperref}
\hypersetup{
 linkcolor=blue,urlcolor=blue,citecolor=blue}
\makeatletter
\providecommand{\tabularnewline}{\\}
\usepackage{blindtext}

\makeatother
\begin{document}
\title{Utilizing anticoincidence veto in a search for gravitational-wave transients}
\author{Souradeep Pal}
\email{sp19rs015@iiserkol.ac.in}

\affiliation{Indian Institute of Science Education and Research Kolkata, Mohanpur,
Nadia - 741246, West Bengal, India}
\begin{abstract}
We devise a technique to suppress the effect of noise transients occurring at gravitational-wave detectors based on temporal anticoincidence. Searches for gravitational-wave signals in the detector data are prone to spurious disturbances of terrestrial origin. The technique presented here benefits from the fact that the noise effects are generally non-coincident in time at geographically separated detectors. Therefore, abnormally loud detector triggers that are not time-coincident can be vetoed. We implement the veto technique in a matched-filter search for transient signals from binary black holes and observe search backgrounds to be generally close to the Gaussian limit. An improvement in the sensitivity of the search is demonstrated using simulated signals. The technique is expected to especially improve the detection efficiency of the search for short duration gravitational waves.
\end{abstract}
\date{\today}
\maketitle
\section{Introduction}\label{sec:intro}
The Advanced LIGO, the Advanced Virgo and KAGRA concluded their fourth observing run (O4) on 18~November 2025. Detection of transient gravitational wave (GW) signals has undergone immense leap, both in quantity and in quality~\citep{PhysRevX.9.031040,PhysRevX.11.021053,PhysRevD.109.022001,PhysRevX.13.041039,ligo2025gwtc,kw5gd732,Abac_2025,twin}. Thanks to the advancements in detector technologies as well as the improvements in detection algorithms~\citep{Aasi_2015,PhysRevD.111.062002,Soni_2025,PhysRevD.102.062003,PhysRevLett.123.231107,PhysRevLett.123.231108,Acernese_2023,Acernese_2015,10.1093/ptep/ptaa125,PhysRevD.111.042010,PhysRevD.93.042004,PhysRevD.105.083018,PhysRevD.108.043004,PhysRevD.109.042008,mbtao4,Dal_Canton_2021,PhysRevD.105.024023}.

The first confident detection of GW signal from a binary black hole (BBH) merger was made in 2015 during the first observing run (O1) of the Advanced LIGO~\citep{abbott2016observation,abbott2016gw150914}. Till the first part of the fourth observing run (O4a) of the GW detector network, observation of more than 200 transient signals have been reported. The observing runs typically last for several months. During an observing run, the GW detectors strive for maximum observing time, however, the observations can be affected due to scheduled maintenance or detector downtimes. A vast majority of the confident detections come from coincident observation of candidate astrophysical signals at geographically separated detectors. The detections are, however, inhibited by the presence of noise artifacts in the detector data.

Searches for the candidate GW signals from the detector data involve multiple steps. In the observing mode, the output of a detector primarily comprises a stationary Gaussian noise floor. It also contains instances of abrupt disturbances of instrumental or environmental origin, local to a given detector. In addition to these terrestrial disturbances, the detector output contains occasional instances of transient astrophysical GW signals. Such signals are expected to be coincident in time across all observing detectors. Therefore, coincident observations generally enhance the chances for confident detection of a GW candidate signal. The presence of noise transients however, degrades the sensitivity of the searches toward the GW transients. The sensitivity of a search can depend upon several factors, including the response of the search to the noise transients.

\begin{figure*}[t]
\begin{centering}
\includegraphics[scale=0.58]{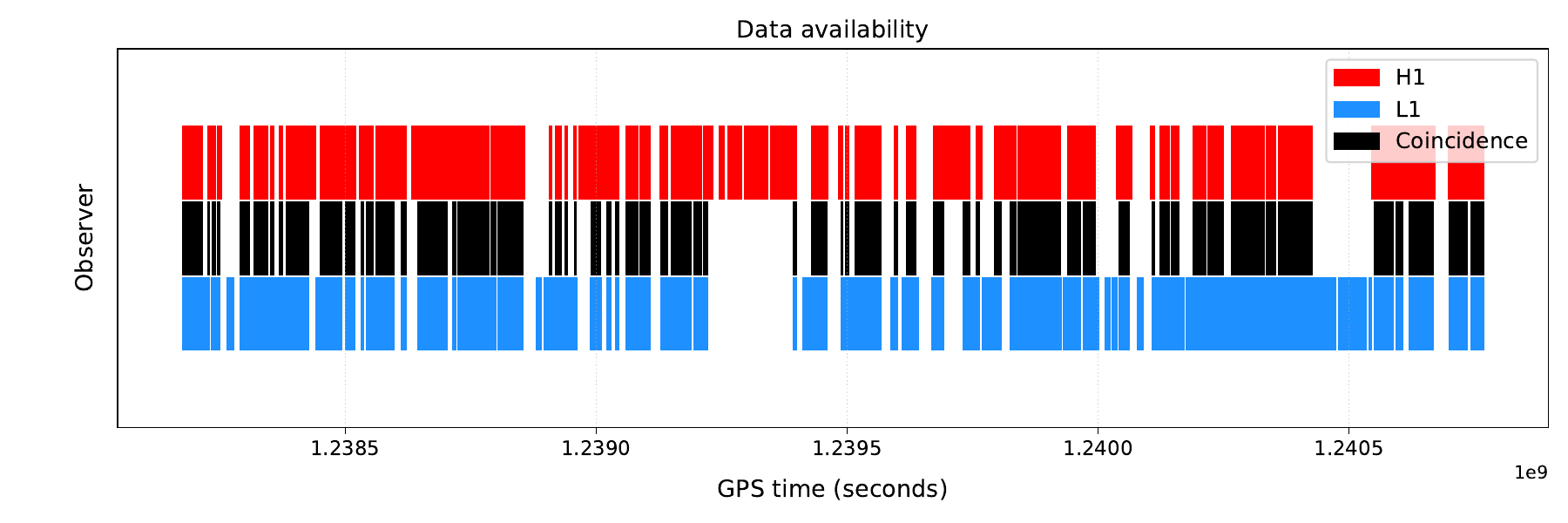}
\par\end{centering}
\caption{Data blocks obtained as coincident observation times between LIGO Hanford (H1) and LIGO Livingston (L1) during their third Observing Run (O3). Here a total of about 30 days of observations is shown which resulted into approximately 14.5 days of coincident observing time.}\label{fig:segs}
\end{figure*}
\begin{figure*}[t]
\begin{centering}
\includegraphics[scale=0.58]{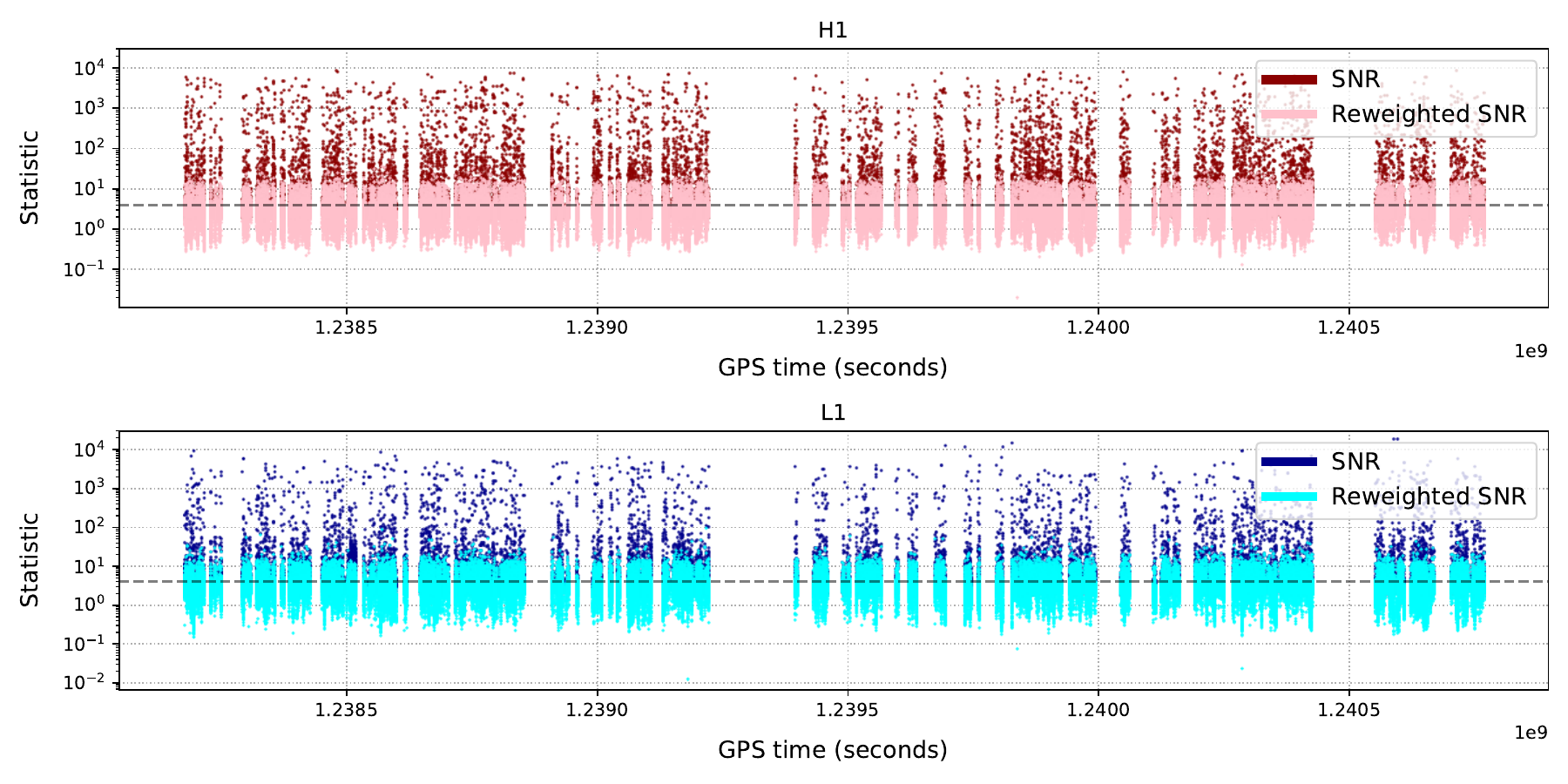}
\par\end{centering}
\caption{Triggers generated from the optimized templates obtained for the coincident data stretches. The SNR and the reweighted SNR for each trigger is plotted against its GPS timestamp. The triggers above the horizontal dashed line in grey (threshold of $\sim$~4.0 on the SNR) are retained for the next stage of the analysis. Also, note that a significant number of triggers can already be discarded by comparing their SNR and reweighted SNR values.}\label{fig:trig}\vspace{-0.19cm}
\end{figure*}
\begin{center}
\begin{table}[b]
\begin{centering}
\begin{tabular}{>{\centering}p{5.0cm}>{\centering}p{3.0cm}}
\hline
{\footnotesize{}Parameter} & {\footnotesize{}Range}\tabularnewline
\hline
\hline
{\footnotesize{}Primary mass, $\mathrm{m_{1}}$ ($\mathrm{M_{\odot}}$)} & {\footnotesize{}[5, 150]}\tabularnewline
{\footnotesize{}Secondary mass, $\mathrm{m_{2}}$ ($\mathrm{M_{\odot}}$)} & {\footnotesize{}[5, 150]}\tabularnewline
{\footnotesize{}Primary aligned-spin, $\mathrm{\chi_{1z}}$} & {\footnotesize{}[-0.99, 0.99]}\tabularnewline
{\footnotesize{}Secondary aligned-spin, $\mathrm{\chi_{2z}}$} & {\footnotesize{}[-0.99, 0.99]}\tabularnewline
\hline
\end{tabular}
\par\end{centering}
\caption{Summary of the target parameter space of the search used to demonstrate the veto. For ease of implementation, we additionally impose $\mathrm{m_{1}}$ $\geq$ $\mathrm{m_{2}}$ during the search.}\label{tab:range}
\end{table}
\par\end{center}
\begin{figure*}[t]
\begin{centering}
\includegraphics[scale=0.58]{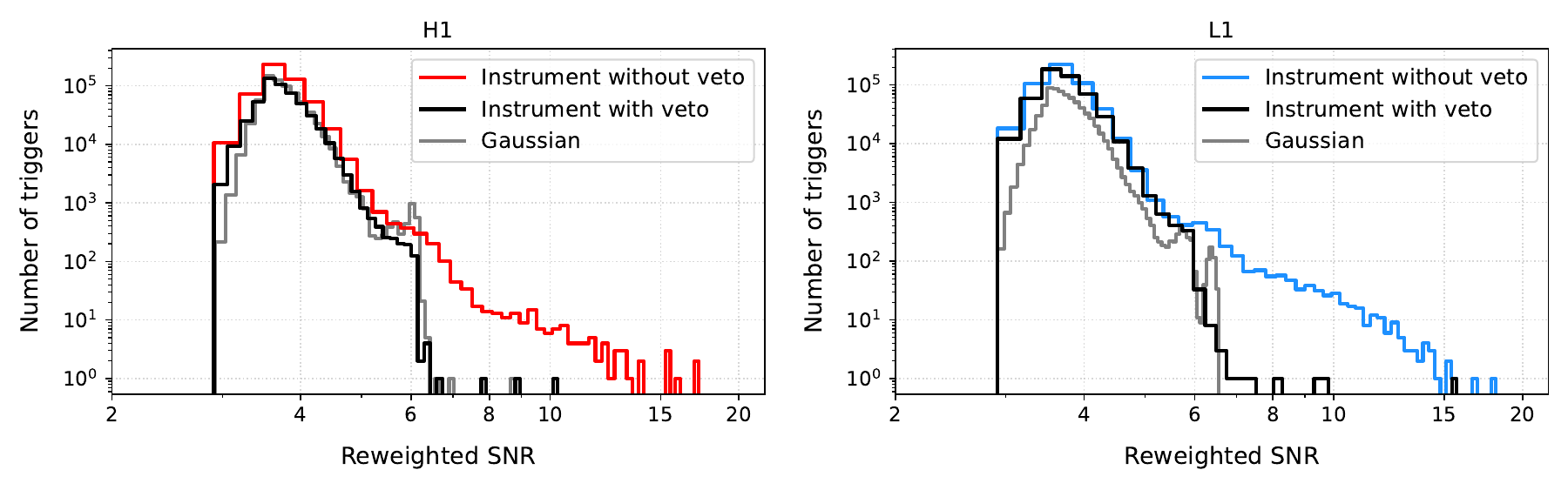}
\par\end{centering}
\caption{Impact of the anticoincidence veto on the distribution of detector triggers ranked by their reweighted SNRs. These exclude those that are caused due to the injections. We observe that the veto step discards a significant number of abnormally loud triggers from each of the detectors, resulting in a distribution which is close to one obtained from Gaussian noise. However, there are very few reasonably loud triggers failing in the anticoincidence test, which could possibly be of astrophysical origin.}\label{fig:impact}\vspace{-0.19cm}
\end{figure*}
\vspace{-1.0cm}

Noise transients can come in a variety of types and morphology. There are several studies that attempt to identify affected times or \textit{glitches} in the instrument data and possibly distinguish them from candidate astrophysical signals~\citep{zevin2017gravity,Vazsonyi_2023,PhysRevD.102.084034,PhysRevD.108.122004,PhysRevD.88.062003,BAHAADINI2018172,mackenzie2025hunting,Ferreira_2025,Wu_2025,Glanzer_2023,Laguarta_2024,PhysRevD.104.102004,chan2025gspynettrees,PhysRevD.97.101501,Davis_2022,PhysRevD.108.124061,PhysRevD.111.103020,PhysRevD.106.023027,Mogushi_2021,PhysRevD.103.044035,dpmn-c6lm,allen2005chi,Nitz_2018,Cabero_2019,Smith_2011}. However, the rate of coincident glitches at various detectors is expected to be small, as reported for the blip glitches in the Advanced LIGO detectors, from the first two observing runs~\citep{Cabero_2019}.

In this work, we implement a veto technique to reduce the effect of noise transients in searches based on temporal \text{\textit{anticoincidence}}. We demonstrate the technique on a search for signals from compact binary mergers. The search relies on the principle of coincident detection, i.e., true astrophysical signals must be coincident in time (and the source properties) across multiple detectors. Conversely, the veto technique utilizes the fact that noise transients are generally non-coincident across any geographically separated detectors, and hence can be rejected or \textit{vetoed}. The search also makes use of the existing vetoing techniques, e.g., those that are based on signal-based consistency tests. The existing techniques generally distinguish signals from glitches at individual detectors based on their time-frequency evolution. The anticoincidence veto appropriately leverages the efficacy of such techniques. We show that it can benefit a search while properly accounting for the vetoed times. We describe the procedure in the search based on particle swarm optimization (PSO) algorithm, previously described in~\citep{pal2023swarm,Pal_2025,PhysRevD.111.104070}. The sensitivity of the search is investigated for the target parameter space(s) that are generally prone to the effect of noise transients, such as those for GW signals from BBHs.

The general techniques used in the search are outlined as follows. The search relies on matched-filtering the data from the detectors~\citep{PhysRevD.44.3819}. It uses a template waveform that models the incoming GW signal, which is generated using the~\texttt{IMRPhenomXAS} model, computed directly in the frequency domain~\citep{PhysRevD.102.064001}. The power spectral density (PSD) computed from the detector data is used which is obtained following Welch's method~\citep{1161901}. A matched-filter results into a signal-to-noise ratio (SNR) timeseries per detector from which \textit{triggers} are generated by thresholding~\citep{usman2016pycbc}. The SNR is reweighted using the power \text{$\mathrm{\chi\textsuperscript{2}}$-veto}~\citep{allen2005chi}, followed by the sine-Gaussian veto applicable for short duration templates~\citep{Nitz_2018}. We select a duration of 30 days during the first half of the third observing run (O3a) of the Advanced LIGO~\citep{abbott2023open}. Simulated signals are injected into the data and the PSO search is used to recover the signals~\citep{pal2023swarm}. To assign a statistical significance to a candidate \textit{event}, the time-sliding approach is used~\citep{wkas2009background}. The search background and other aspects of the search are investigated with and without using the anticoincidence veto introduced in this work.

The article is organized into the following sections. Section~\ref{sec:sec1} describes the process used to set up the search. We generally skip the detailed description of the computational steps since they are reported elsewhere as mentioned earlier. Section~\ref{sec:sec2} describes the implementation of the anticoincidence veto in the search. Comparisons of several aspects of the search with and without the veto are included. Section~\ref{sec:sec3} describes an assessment of the detection efficiency of the search using the injected signals. A few aspects of the search background involving higher-mass systems and higher rate of signal detection are discussed in the appendix sections.
\vspace{-0.4cm}
\section{Search description}\label{sec:sec1}
In this section, we describe an overview of the steps associated with the search. The preparatory steps to organize the detector data for the analysis are described as follows. The duration between a certain GPS start and end time is considered for the analysis. Times are identified where coincident data are available for at least two detectors. For simplicity of demonstration, data from LIGO Hanford (H1) and LIGO Livingston (L1) are considered. This is illustrated in Fig.~\ref{fig:segs}. A single continuous stream of data is referred to as a data \textit{block}. The coincident data blocks are assumed to be present on computer storage before the start of the analysis. The analysis does not explicitly require any data quality information.

Any coincident data block is divided into multiple smaller overlapping \textit{stretches} of $T$~$\sim$~128 seconds duration wherever possible. The overlap of each stretch with its neighbouring stretches is by a total of 50\% of the duration $T$. In the case of unavailability of a sufficient duration in a neighbouring stretch or the given stretch itself is shorter than $T$ seconds, an appropriate duration of coincident data from the next neighbouring block(s) is augmented. This is to ensure uniformity in the duration of the data stretches, however, triggers from only the original duration $T$ are considered. The triggers are generated from matched-filtering the detector data for the individual data stretches using the PSO algorithm, briefly described as follows.
\begin{table}[b]
\centering
\begingroup
\begin{center}
\begin{circuitikz}[scale=5]
\ctikzset{bipoles/length=2.8cm}
\draw (0,0) node[xor port] (myxnor) {};
\draw (-0.225,0) node {XOR};
\draw (-0.6,0.2) node {A};
\draw (-0.6,-0.2) node {B};
\draw (0.05,0.1) node {Y};
\end{circuitikz}
\end{center}
\endgroup
\vspace{0.02cm}
\begin{center}
\begin{centering}
\begin{tabular}{>{\centering}p{2.5cm}>{\centering}p{1.8cm}>{\centering}p{1.8cm}>{\centering}p{1.8cm}}
\hline 
{\footnotesize{}\makecell{Input A \\ (H1 trigger)}} & {\footnotesize{}\makecell{Input B \\ (L1 trigger)}} & {\footnotesize{}\makecell{Output Y \\ (Veto)}}\tabularnewline
\hline 
\hline 
{\footnotesize{}0} & {\footnotesize{}0} & {\footnotesize{}0}\tabularnewline
{\footnotesize{}1} & {\footnotesize{}0} & {\footnotesize{}1}\tabularnewline
{\footnotesize{}0} & {\footnotesize{}1} & {\footnotesize{}1}\tabularnewline
{\footnotesize{}1} & {\footnotesize{}1} & {\footnotesize{}0}\tabularnewline
\hline 
\end{tabular}
\end{centering}
\caption{The exclusive~OR (XOR) logic is used to veto unusually loud anticoincident triggers that survive beyond the standard signal-based consistency tests.}\label{tab:logic}
\end{center}
%\vspace*{2\floatsep}
\end{table}
\begin{figure*}[t]
\begin{centering}
\includegraphics[scale=0.70]{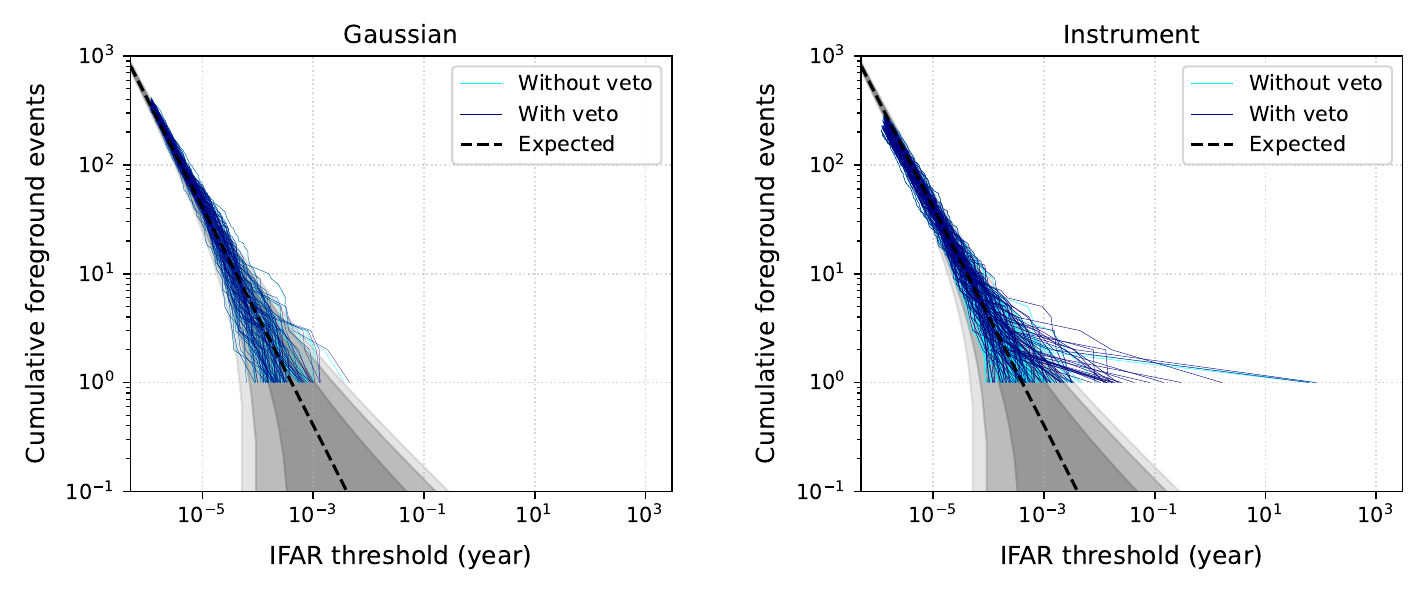}
\par\end{centering}
\caption{Self-consistency of the FAR estimates with and without the anticoincidence veto in Gaussian noise (left) and in instrument noise (right) from O3a. The effect of the veto is demonstrated on the same corresponding datasets split into smaller chunks using identical analysis configuration.}\label{fig:fars}\vspace{-0.19cm}
\end{figure*}

The parameter space for the search is indicated in \text{Table~\ref{tab:range}}. In the PSO algorithm, template points are dynamically placed in the search parameter space. For any given stretch of data, the algorithm optimizes a \textit{ranking} (or detection) statistic over the search space \text{in iterations}. The ranking statistic here is the quadrature sum of the reweighted SNRs computed over the given detector network. To compute the statistic for a given template point, triggers are generated by matched-filtering the data from individual detectors and picking the instances whenever the resulting SNR timeseries crosses a threshold of $\sim$~4. The SNR values are then reweighted using the power $\mathrm{\chi\textsuperscript{2}}$-veto and the sine-Gaussian veto, wherever applicable. Triggers with the ratio of reweighted SNR to SNR smaller than unity (typically below 0.75) are discarded. The surviving triggers are ranked by their reweighted SNR values and the ranking statistic is computed for the given template. To optimize the statistic for the data stretch, template points are iteratively regenerated using the PSO algorithm until a fixed number of iterations is reached. In total, approximately 9000~template points are computed per data stretch, which are sufficient for optimizing the ranking statistic on average. Note that the template points and the triggers described above are primary in nature and are only used to obtain the \textit{optimized} template for the data stretch and therefore, are not retained in the analysis further.

Here we continue by assuming that the algorithm furnishes an optimized template point as the output from any given coincident data stretch. The secondary triggers, which are computed from the optimized template in the same fashion as described above, undergo the same veto processes as the primary triggers. The collection of these triggers over all data stretches is depicted in Fig.~\ref{fig:trig}. These are additionally tested for time-coincidence (vetoed and \textit{foreground} events generated) and are used to assign a statistical significance to the generated foreground events, as described below. Here a 15~ms window is used for the HL detector network, given the travel time of light between the two detectors. Sufficiently loud triggers of astrophysical origin are expected to pass the time-coincidence test and feature as foreground events, whereas any non-Gaussian triggers of terrestrial origin are likely to pass a time-anticoincidence test as described in the next section. Note that the only point of contact between the data and the search is through the optimized templates obtained from the data stretches.

Whenever a foreground candidate is observed in the triggers from the individual detectors, a \textit{buffer} of sufficient duration is formed from the triggers available immediately before the candidate. Note that the foreground events are obtained from triggers coming from a fixed (the optimized) template in a given data stretch, which ensures the source-parameter consistency across the detectors, as is expected in the case for astrophysical signals. In contrast, \textit{background} events, as described below, are obtained from buffering the triggers over multiple data stretches that need not maintain this consistency. Since noise triggers in a given detector can arise from any part of the parameter space, the collection of such triggers are thus not expected to have the source-parameter consistency.

The buffer duration (of typically several hours) is preset depending upon the upper limit of the significance desired. For the significance estimation, triggers are treated in a continuous (and cyclic) fashion from the coincident data blocks. Therefore, the triggers from two adjacent coincident data blocks, but possibly separated by (even) a large time interval, are used in buffering triggers in order to assign significance to a given foreground candidate. Specifically, this is a reasonable assumption given that the veto process effectively minimizes the non-Gaussianity and the non-stationarity of the noise (triggers) from the instruments, as shown in Fig.~\ref{fig:impact} and discussed further in the next section.

\begin{figure*}[t]
\begin{centering}
\includegraphics[scale=0.70]{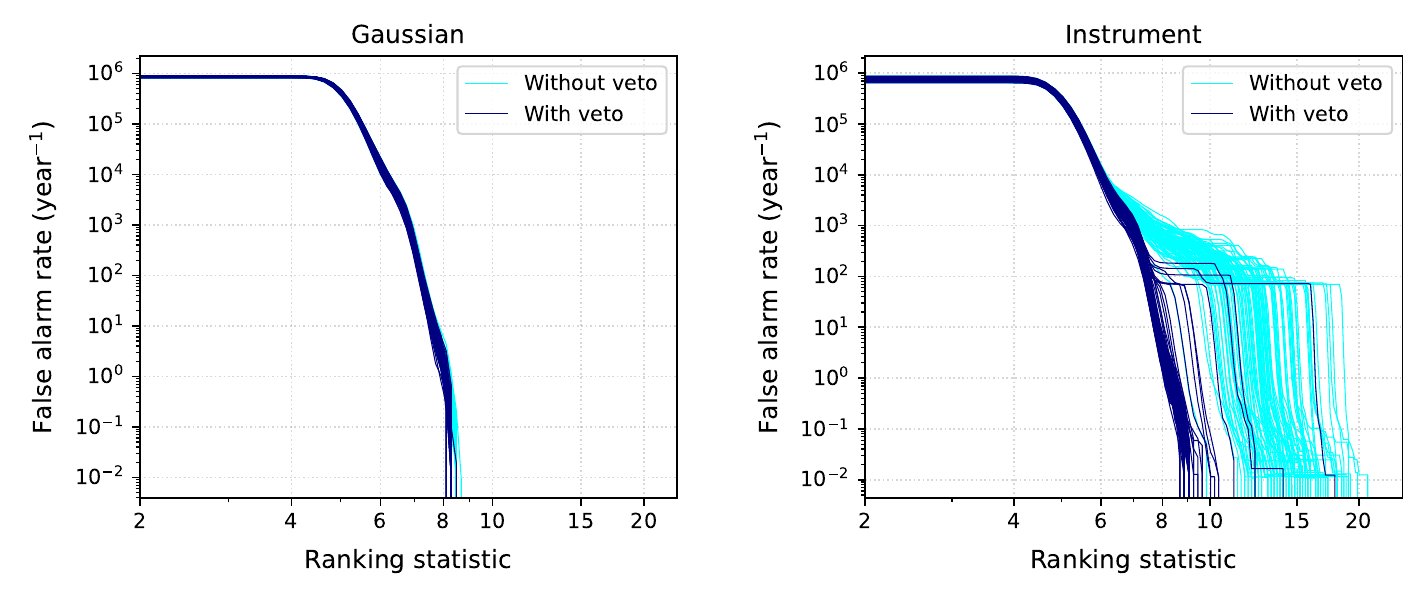}
\par\end{centering}
\caption{Distribution of background events over ranking statistic with and without the anticoincidence veto applied. We observe that while in Gaussian noise (left), the veto has negligible effect on the distributions, the same has a large effect in generally obtaining Gaussian-like distributions in instrument noise (right).}\label{fig:rocs}\vspace{-0.19cm}
\end{figure*}
The triggers obtained during the buffer time from the detectors are artificially time-slided by an interval greater than the light travel time between the detectors. An unphysical time-slide duration~(of~$\sim$~50~ms) ensures that any coincidence between triggers hence formed are non-astrophysical in nature. The false alarm rate (FAR, or equivalently the significance or the inverse FAR) is assigned to a foreground candidate based on the number of such background events having a ranking statistic larger than the given foreground event. A large sample of background events is generated with as many time-slides as possible, which is limited by the number of distinct time-slides possible with a given buffer duration. When the procedure described above assigns acceptable FAR values to foreground events and an appropriate number of foreground events are generated using the triggering process described earlier, we observe an empirical self-consistency, as shown in Fig.~\ref{fig:fars}.

In O3-like scenarios, the entire analysis requires modest computing resources ($\sim$~8~CPU cores and $\sim$~8~GB memory) to finish the computation for a given stretch of coincident data in time before the next stretch is analyzed. Triggers from only the optimized templates are stored on the computer disk, which occupy $\sim$~10~GB for a full typical observing run.
\vspace{-0.4cm}
\section{Anticoincidence veto}\label{sec:sec2}
In the previous section, we discussed the process of generation of triggers from individual detectors and the application a ratio test based on the SNR and the reweighted SNR values of a trigger, computed based on the power $\mathrm{\chi\textsuperscript{2}}$-veto and the sine-Gaussian veto (wherever applicable). For the surviving triggers, the distribution of the reweighted SNRs for the individual detectors are shown in Fig.~\ref{fig:impact}. The same is indicated when obtained by replacing the datasets with Gaussian noise instead of instrument noise, but keeping the detector PSDs identical. We observe that for any detector, the distribution from the instrument noise contains an excess of triggers as compared to that from the Gaussian noise. However, note that the reweighted SNR values of such triggers are generally small (typically $<$~20), and hence, a majority of these are not likely of astrophysical origin.

An astrophysical trigger in any detector should be accompanied with a coincident counterpart during a coincident observing time\footnote{There are exceptions to this, e.g., when the source is located around a \textit{blind spot} of a detector arising due to antenna pattern functions, or due to a reduced sensitivity of the detector during observations; such cases are expected to be small in number.}. Therefore, the triggers in excess can be effectively vetoed. We devise a simple technique to veto such triggers based on anticoincidence. Triggers in a given detector that are louder than a certain threshold (of $\sim$~6 in the reweighted SNR) are tested for time-anticoincidence with any nearby trigger from the remaining detector(s). If the test is positive, the veto is turned on and the triggers are removed, effectively \textit{masking} the trigger streams for a short duration as described below. When the test is negative, the triggers from each detector are retained. A summary of this logic is shown in Table~\ref{tab:logic}. We observe that the distribution of the resulting triggers after the veto process is close to the Gaussian case in both the detectors. Note that there are a few loud surviving triggers in the detectors which fail the anticoincidence test, i.e., they pass the coincidence test and can be potentially astrophysical in origin. Broadly, the veto technique thus reduces the non-Gaussian and non-stationary component of the instrument data.

The times when the veto is turned on, any trigger during a small period around the vetoed time is removed from each detector. The procedure is equivalent to masking the datasets during these periods as if they are effectively not observed. As these instances are generally small in number, they are not expected to affect the search in a significant way. In the 30 days of observations considered, with $\sim$~14.5 days of coincident time, triggers for around 30 minutes of the coincident data are vetoed on aggregate, following this procedure on both the detectors. Here, the triggers are originally generated at the rate of no more than once per second. If triggers are generated on a shorter time scale, the aggregate duration of such masked times can be reduced\footnote{Note that in practice, the trigger stream is masked momentarily and not the data stream itself (and hence the term `veto' may be more appropriate rather than a `gate').}. Since the vetoed times are thus accounted for, we do not observe the veto process to interfere with the FAR estimates as shown in Fig.~\ref{fig:fars}. We also observe that the distribution of the background events obtained with the veto is generally close to that in the Gaussian limit, as shown in Fig.~\ref{fig:rocs}. This has a direct consequence in the recovery of signals from instrument data, which is described in the next section.

Since the time-coincidence test requires a trivial computation, the overall cost for the application of the veto is negligibly small. For instance, this one-time step can be completed within a few seconds on a single CPU for the entirety of a typical observing run.
\vspace{-0.3cm}
\section{Detection efficiency}\label{sec:sec3}
In this section, we demonstrate the impact of the veto technique on the sensitivity of the search. To evaluate the sensitivity, simulated signals are injected into real instrument data and are recovered by the search with and without the veto applied. The signal parameters come from the injection sets containing a representative population of the astrophysical sources~\citep{o3popzenodo}. The injections are pre-selected from the coincident duration used earlier, and are bounded by the signal parameter ranges given in Table~\ref{tab:range}. In addition to the aligned-spin components, the injections also carry precessing spins, implemented using the \texttt{IMRPhenomXP} waveform approximant~\citep{PhysRevD.103.104056}.

\begin{figure}[t]
\begin{centering}
\includegraphics[scale=0.66]{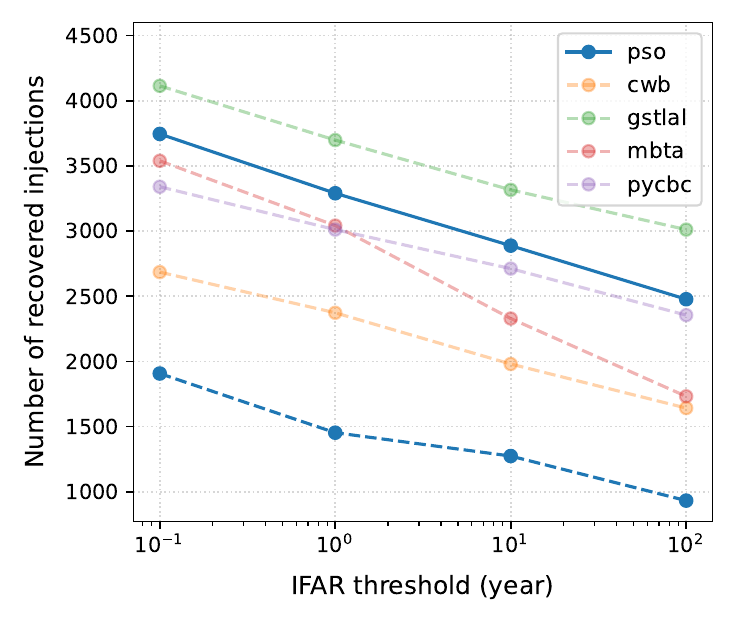}
\par\end{centering}
\caption{Comparison of the number of injections recovered with (solid line) and without (dashed line) using anticoincidence veto from the same datasets. For reference, the injection recovery rates of the analyses participating in O3 are shown~\citep{o3popzenodo}. These indicate injections recovered at the given IFAR thresholds from the same coincident times.}\label{fig:sens}\vspace{-0.19cm}
\end{figure}

A search is conducted on the triggers, including those due to the injections, using the same analysis configuration as described in the earlier sections. Similar to the analysis framework, the triggers from both detectors are channeled through two different processes- with and without the veto step applied. For each process, foreground candidates are computed from the triggers and a FAR estimate is assigned to each candidate. Triggers around the injection times are properly accounted for in order to prevent them from affecting the FAR estimates of any foreground candidate. The injections are identified from the foreground events as follows. An injection is considered to be \textit{recovered} if there exists a foreground candidate with triggers at the two detectors appearing in coincidence within 1 second around the injected coalescence time above a significance threshold. The number of the recovered injections up to a given significance threshold (IFAR) with and without the veto step applied is summarized in Fig.~\ref{fig:sens}. We observe that the number injections recovered with the veto applied is greater than that without the veto, at any of the given IFAR thresholds. This is attributed to the improved backgrounds with the veto as indicated in Fig.~\ref{fig:rocs}.

\begin{figure}[t]
\begin{centering}
\includegraphics[scale=0.7]{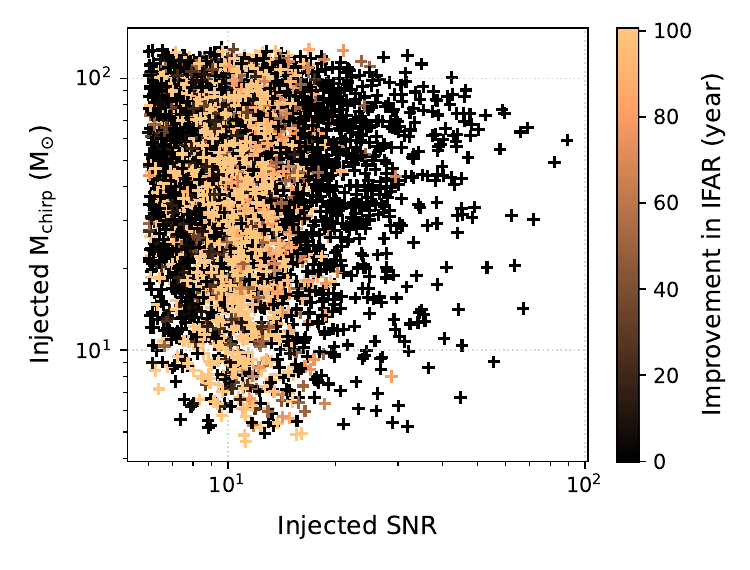}
\par\end{centering}
\caption{Summary of the improvement in the IFAR estimates in the recovered injections. We observe generally greater improvements for injections with moderate SNRs. The injections that are missed in the search are not investigated here.}\label{fig:reco}\vspace{-0.19cm}
\end{figure}

Here the recovery of injections with small to moderately large ranking statistic (or loudness in general) are expected to benefit more from the veto process. This is summarized in Fig.~\ref{fig:reco}. The injections with very small SNRs (say $\le$~8) generally have small significance estimates- at these signal strengths, a large number of louder background events can be expected due to triggers of Gaussian noise origin, largely deciding their significance estimates. On the other hand, very loud signals (with SNRs $\ge$~20) are not expected to observe much improvement since they already outstand the majority, if not all  of the background events. From these injections, we do not apparently observe any strong dependence of the improvement on the chirp mass ($\mathrm{M_{chirp}}$) of the systems.

We note that in the search, 128-seconds segments are used for recovering the injections. If the rate of the injections is large such that there are multiple injections within a given data stretch, the PSO algorithm can confuse itself and therefore cannot optimize the source parameters for these joint injections simultaneously. Considering injections that are above a fiducial network SNR of 6, the abundance of such joint injections in the current case is around 35\%. This leads to a reduced sensitivity toward such injections\footnote{We do not quantitatively assess this effect here, however, the rate of injections is expected to be greater in subsequent observing runs, e.g., in O4a~\citep{o4asens}.}. The rate of detectable astrophysical signals is much smaller. Thus, the sensitivity of the PSO search reported could be slightly underestimated, i.e., the true sensitivity in a real scenario is expected to be somewhat higher; refer to Appendix~\ref{sec:appn1}. This is different from the issue with a larger rate of astrophysical signals observable in the subsequent observing runs, which needs to be tackled in the future; refer to Appendix~\ref{sec:appn2} for a related discussion. Even a small number of overlapping signals may be expected because of the improvement targets in the sensitivity of the detectors.
\vspace{-0.3cm}
\section{Conclusions}\label{sec:concl}
\begin{figure*}[t]
\begin{centering}
\includegraphics[scale=0.70]{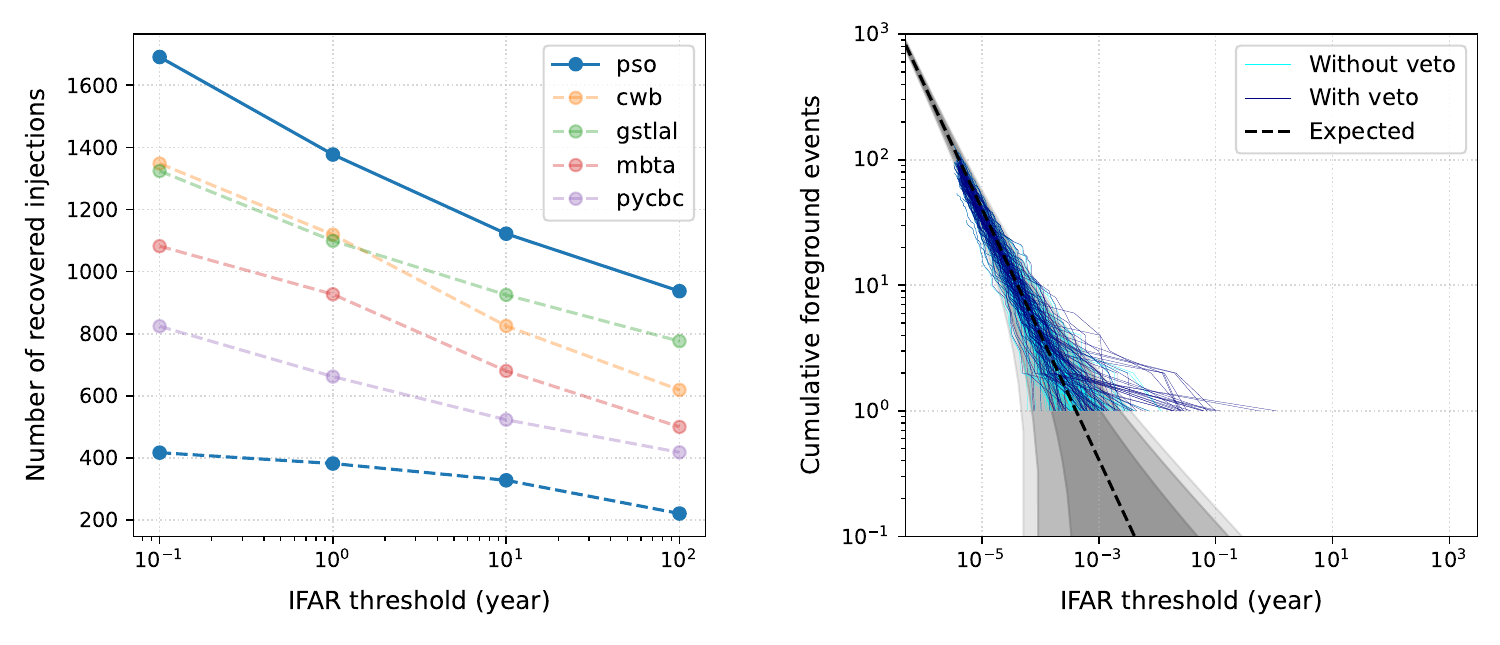}
\par\end{centering}
\caption{Number of high-mass injections recovered at different significance thresholds (left) and the self-consistency of the significance estimates (right) obtained with and without using anticoincidence veto in O3a data.}\label{fig:imbh}\vspace{-0.25cm}
\end{figure*}
Searches for gravitational-wave transient signals are prone to the effects of noise transients in the data from gravitational-wave observatories. The noise transients, mainly in the form of glitches, inhibit the sensitivity of the searches, especially for short duration signals. We have explored a technique to reduce the effect of the noise transients using data from multiple observatories. The technique utilizes the temporal anticoincidence in the occurrence of the noise transients at spatially separated detectors. It is effective regardless of the type of the glitch present in data, concisely bypassing the need to identify, model, and subtract them. However, the benefit can possibly be compromised in any rare case of time-coincident glitches. Single detector observation times are not eligible for the technique.

We demonstrate that the veto process reduces the non-stationarity and the non-Gaussianity of the instrument data, improving the search backgrounds from the perspective of a modeled gravitational-wave search. In principle, the veto can benefit any search that utilizes a coincident detection strategy irrespective of being modeled or not.

The technique has been implemented in a search targeting spin-aligned binary black hole sources. The improvement in the sensitivity of the search is demonstrated using simulated signals. We find that the detection efficiency improves with the use of the veto process. We have also pointed out an origin of possible underestimation of the sensitivity reported from the PSO search. The sensitivity can be small for the injected signals which occur together in any given analysis stretch. Appropriate workarounds may be adopted in the future, such as by performing multiple rounds of the analysis from the same injection set but requiring at most a single injection per data stretch in any given round. This is different from the future scenario where we expect loud (or overlapping) astrophysical signals observable at a greater rate. In such cases, the significance estimated readily from the time-shifting procedure could be biased. Nevertheless, the recovery of injections using the search in the current scenario is broadly comparable to some of the existing searches. The sensitivity of the search can be further improved by making use of the phase and the amplitude (or the SNR) information of potential astrophysical signals, see for example~\citep{Nitz_2017,davies2020extending,PhysRevD.104.042004}. Further, the search can take into account the population of the sources being detected to improve its sensitivity, see for example~\citep{PhysRevD.110.043036}.

Here we have used only two detectors for demonstration and have not considered any possible single-detector events in the coincident times. A search including single detector events can be explored in the future which could utilize the byproduct triggers of this search. On the other hand, the technique can also be extended to a larger network of detectors especially with comparable sensitivities.
\vspace{-0.3cm}
\section{Acknowledgments}\label{sec:ackno}
This research has made use of data or software obtained from the Gravitational Wave Open Science Center (gwosc.org), a service of the LIGO Scientific Collaboration, the Virgo Collaboration, and KAGRA. This material is based upon work supported by NSF's LIGO Laboratory which is a major facility fully funded by the National Science Foundation, as well as the Science and Technology Facilities Council (STFC) of the United Kingdom, the Max-Planck-Society (MPS), and the State of Niedersachsen/Germany for support of the construction of Advanced LIGO and construction and operation of the GEO600 detector. Additional support for Advanced LIGO was provided by the Australian Research Council. Virgo is funded, through the European Gravitational Observatory (EGO), by the French Centre National de Recherche Scientifique (CNRS), the Italian Istituto Nazionale di Fisica Nucleare (INFN) and the Dutch Nikhef, with contributions by institutions from Belgium, Germany, Greece, Hungary, Ireland, Japan, Monaco, Poland, Portugal, Spain. KAGRA is supported by Ministry of Education, Culture, Sports, Science and Technology (MEXT), Japan Society for the Promotion of Science (JSPS) in Japan; National Research Foundation (NRF) and Ministry of Science and ICT (MSIT) in Korea; Academia Sinica (AS) and National Science and Technology Council (NSTC) in Taiwan.

The research and development was carried out at the Center of Excellence in Space Sciences India (CESSI). CESSI is a multi-institutional Center of Excellence hosted by the Indian Institute of Science Education and Research (IISER) Kolkata and has been established through funding from the Ministry of Education, Government of India. We acknowledge the use of IUCAA LDG cluster Sarathi for the computational/numerical work. Several tools from the PyCBC software package are used in the analyses. We acknowledge interaction and feedback from several members of the GW community. The author thanks K. Rajesh Nayak for the discussions during this work. The author thanks Johann Fernandes for carefully reading the manuscript and for providing feedback on it. This material has a LIGO document number, P2500796. The author did not receive funding or financial support from any agency or institution.
\vspace{-0.4cm}
\section{Data availability}\label{sec:datav}
The data that support the findings of this article are openly available~\citep{RICHABBOTT2021100658,abbott2023open,gwosc}.
\appendix
\vspace{-0.3cm}
\section{Searches for higher-mass systems}\label{sec:appn1}
Dedicated searches for intermediate mass black hole binaries have been previously carried out~\citep{PhysRevD.100.064064,abbott2022search,PhysRevD.104.042004}. In this section, we implement a search similar to the one indicated in Table~\ref{tab:range}, except that the targeted component masses now belong to $[100, 250]$~$\mathrm{M_{\odot}}$. This further exposes the search to the possible glitches triggered by relatively higher-mass (shorter duration) templates. The sensitivity of a search can be optimized by using, for example, a strict cut on the duration of templates~\citep{PhysRevD.104.042004}. Here we skip such fine optimization choices, primarily to isolate the effectiveness of the veto step.

The injections within the given mass range are recovered from a longer duration of coincident data during O3a in a similar fashion as described in the main text. These are summarized in Fig.~\ref{fig:imbh}. We observe that the number of injections recovered is greater while using the anticoincidence veto at any given significance threshold. The self-consistency of the significance estimates is also shown. We attribute the greater improvement in the significance estimates, in part, to the smaller effect of the joint injections within any analysis segment as described in Section~\ref{sec:sec3}, which is around 4\% here due to a smaller rate of the high-mass injections in the given dataset, with the same analysis configuration. We note that the computational cost of conducting such a search is trivially small.
\vspace{-0.5cm}
\section{Significance estimates in the presence of nearby coincident events}\label{sec:appn2}
\begin{figure}[t]
\begin{centering}
\includegraphics[scale=0.7]{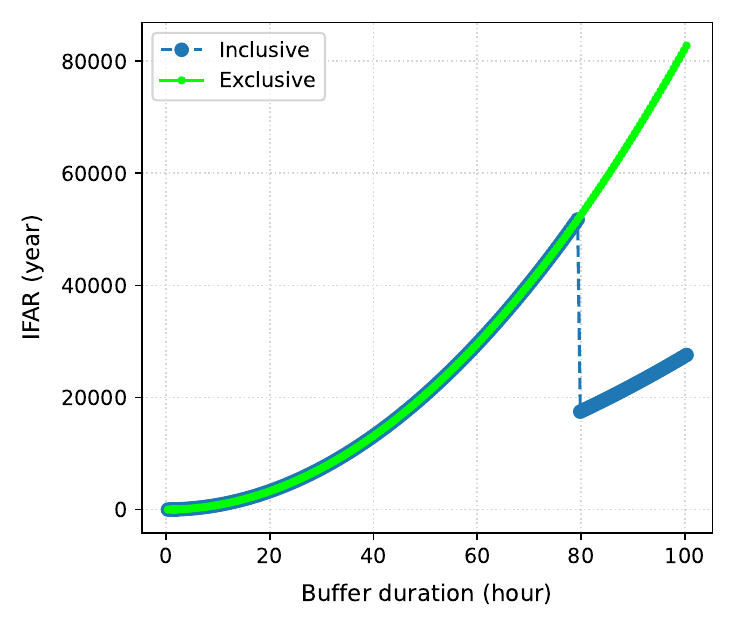}
\par\end{centering}
\caption{Impact on the significance estimate of \texttt{GW190521} due to a nearby coincident candidate in the time-slide approach. The estimate is empirically obtained utilizing the anticoincidence veto, inclusive or exclusive of the nearby coincident candidates from the same search obtaining the given event.}\label{fig:neig}\vspace{-0.39cm}
\end{figure}
In the earlier sections, we discussed the effect of anticoincident triggers on searches for transient signals. However, nearby (loud) coincident foreground triggers of noise origin or of true astrophysical origin, to a given candidate can still limit its significance estimate. This is illustrated for \texttt{GW190521\_030229} (referred to as \texttt{GW190521}) in Fig.~\ref{fig:neig}. The kink in the inclusive significance estimate is due to a coincident candidate at the GPS time~1241858977.0 seconds, around 80 hours backward in coincident time from the event, obtained from the same search. This is a marginal candidate with an IFAR of $\sim$~1.0 year and with a network SNR of 10.65, which inhibits the significance estimate beyond a certain value of $\sim$~50000 years. The coincident events obtained from a few blocks of data around \texttt{GW190521} are summarized in Table~\ref{tab:event}. Here we do not observe such an effect from the confident event~\texttt{GW190519\_153544} (referred to as \texttt{GW190519}), the nearest neighbour to \texttt{GW190521} backward in time, and other remaining candidates, due to their relatively smaller detector-wise ranking statistics. Although very high significance is typically not necessary for the observation of astrophysical candidates, cases such as the one described above are expected to be critical in the upcoming observations given a higher projected rate of detections. To take this into account, an appropriate workaround to the time-slide approach may be needed in the future.
\vspace{-0.9cm}
\begin{center}
\begin{table}[b]
\begin{centering}
\begin{tabular}{>{\centering}p{1.75cm}>{\centering}p{2.0cm}>{\centering}p{1.4cm}>{\centering}p{1.2cm}>{\centering}p{1.8cm}}
\hline
{\footnotesize{}Event name/type} & {\footnotesize{}GPS time (seconds)} & {\footnotesize{}IFAR (year)} & {\footnotesize{}Network SNR} & {\footnotesize{}Reweighted SNR}\tabularnewline
\hline
\hline
{\footnotesize{}GW190521} & {\footnotesize{}1242442967.4} & {\footnotesize{}27397.2} & {\footnotesize{}14.36} & {\footnotesize{}13.01}\tabularnewline
{\footnotesize{}GW190519} & {\footnotesize{}1242315362.3} & {\footnotesize{}20547.9} & {\footnotesize{}12.44} & {\footnotesize{}11.87}\tabularnewline
{\footnotesize{}Marginal} & {\footnotesize{}1241613498.4} & {\footnotesize{}10.2} & {\footnotesize{}11.35} & {\footnotesize{}10.91}\tabularnewline
{\footnotesize{}Marginal} & {\footnotesize{}1242411983.5} & {\footnotesize{}1.1} & {\footnotesize{}9.35} & {\footnotesize{}9.35}\tabularnewline
{\footnotesize{}Marginal} & {\footnotesize{}1241858977.0} & {\footnotesize{}1.0} & {\footnotesize{}10.65} & {\footnotesize{}10.58}\tabularnewline
\hline
\end{tabular}
\par\end{centering}
\caption{Summary of the candidate events from the search. Reported IFARs are all-inclusive. Here events with an IFAR less than 100 years and more than or equal to 1 year are typed marginal. Not all marginal events may be reported elsewhere.}\label{tab:event}
\end{table}
\par\end{center}
\bibliography{veto}
\bibliographystyle{apsrev4-1}
\end{document}